\documentclass[5p]{elsarticle}
\usepackage{url}

\AtBeginDocument{%
  }

\makeatletter
\def\ps@pprintTitle{%
   \let\@oddhead\@empty  % clear header
   \let\@evenhead\@empty % clear header
   \def\@oddfoot{}       % clear footer
   \let\@evenfoot\@oddfoot % clear footer
}
\makeatother

\begin{document}

%%
%% The "title" command has an optional parameter,
%% allowing the author to define a "short title" to be used in page headers.
% \title{Estuary: A Multimodal Framework For Building Low-Latency Real-Time Socially Interactive Agents}

\title{Estuary: A Framework For Building Multimodal Low-Latency Real-Time Socially Interactive Agents}

%%
%% The "author" command and its associated commands are used to define
%% the authors and their affiliations.
%% Of note is the shared affiliation of the first two authors, and the
%% "authornote" and "authornotemark" commands
%% used to denote shared contribution to the research.
\author{Spencer Lin\textsuperscript{*}}
% \ead{linspenc@usc.edu}

\author{Basem Rizk\textsuperscript{*}}
% \ead{brizk@usc.edu}

\author{Miru Jun\textsuperscript{*}}
% \ead{mdjun@usc.edu}

\author{Andy Artze}
% \ead{artze@usc.edu}

\author{Caitl\'in Sullivan}
% \ead{ccsulivan@usc.edu}

\author{Sharon Mozgai}
% \ead{mozgai@ict.usc.edu}

\author{Scott Fisher}
% \ead{scott.fisher@usc.edu}

% Affiliations
\address{University of Southern California}

\tnotetext[preprint]{Denotes equal contribution}
\tnotetext[preprint]{This manuscript is a pre-print version of a paper accepted for publication at the ACM Intelligent Virtual Agents (IVA) 2024 Conference [DOI: 10.1145/3652988.3696198] [ACM ISBN: 979-8-4007-0625-7/24/09].}

%%
%% By default, the full list of authors will be used in the page
%% headers. Often, this list is too long, and will overlap
%% other information printed in the page headers. This command allows
%% the author to define a more concise list
%% of authors' names for this purpose.
% TODO rewrite this
% \renewcommand{\shortauthors}{Doe et al.} 

%%
%% The abstract is a short summary of the work to be presented in the
%% article.
% \onecolumn
\begin{abstract}
 % In recent years, advancements in generative artificial intelligence has had a major impact in nearly every industry. The rapid development and implementation of this technology in all forms has spurred their application to the field of Socially Interactive Agents (SIAs). 
 The rise in capability and ubiquity of generative artificial intelligence (AI) technologies has enabled its application to the field of Socially Interactive Agents (SIAs). Despite rising interest in modern AI-powered components used for real-time SIA research, substantial friction remains due to the absence of a standardized and universal SIA framework. To target this absence, we developed Estuary: a multimodal (text, audio, and soon video) framework which facilitates the development of low-latency, real-time SIAs. Estuary seeks to reduce repeat work between studies and to provide a flexible platform that can be run entirely off-cloud to maximize configurability, controllability, reproducibility of studies, and speed of agent response times. We are able to do this by constructing a robust multimodal framework which incorporates current and future components seamlessly into a modular and interoperable architecture.

\end{abstract}

\maketitle

% Instructions from https://iva.acm.org/2024/call-for-demos/
% Demo paper submissions should be maximum 2 pages long with 1 additional page for references and 1 page detailing all technical requirements for the presentation of the demo. The authors should include in their submissions:

% Demo title
% Authors and Presenters: affiliation, email address, selected contact person
% Rationale: Explain why the system demonstrated is of interest to the IVA research community.
% System explanation: Give a short explanation of the system, explain its use cases and how it helps bridge a specific knowledge gap.
% Video material: We recommend that demo submissions include a hyperlink to a video of the demo. The video should be no more than 5 minutes long.
% The technical requirements page for the demo should include everything necessary to run and display the demo (e.g., one external monitor, two headsets, one table, etc.).
 \begin{figure}
      \centering
      \includegraphics[width=0.5\textwidth]{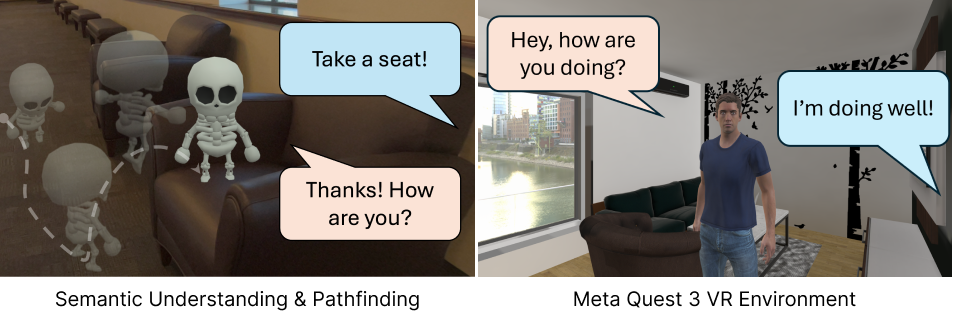}
      % \vspace{-14px}
      \caption{Estuary's various capabilities in Augmented Reality and Virtual Reality environments.}
      \label{fig:teaser}
\end{figure}
% ======================================================

% \vspace{-5px}
\raggedbottom
\section{Background}
Rapid advancements in AI have catalyzed the development of SIAs, which integrate complex technologies to facilitate nuanced human-computer interactions across various domains, producing affective virtual agents~\cite{hasan2023sapien}, multi-agent simulations of human behavior~\cite{park2023generative, wang2023humanoid, gao2023s}, and more~\cite{lieconagent, liang2023encouraging, shaikh2024rehearsal}. To build effective SIAs, multiple microservices and components need to be integrated and managed, including Automatic Speech Recognition (ASR) and Text-To-Speech (TTS). Implementing these features require significant efforts. A modern, comprehensive SIA framework is useful to streamline efforts and reduce redundant work~\cite{lugrin2022handbook} and to provide a scalable and interoperable standard that can support new models and microservices.

% where SIA research is going: LLMS, deep learning voice 
% \textbf{LLMs and SIAs}
% Other research has been towards creating social interaction datasets~\cite{kokomind2023} and social intelligence and reasoning benchmarks~\cite{zhou2023sotopia, gandhi2024understanding} for driving quality LLM research. However, there is a gap to be filled: the fundamental need for a flexible and intuitive framework for building LLM-powered SIAs to expedite development.

% - general virtual agent: https://arxiv.org/pdf/2308.03022
% - but many are constrained to game environments 
% - generative agents, specifically LLMs

% --------------------

% related works
\subsection{SIA \& Conversational AI Frameworks.}
Several existing toolkits help streamline the process of building SIAs. The Virtual Human Toolkit (VHToolkit)~\cite{hartholt2013all} and Greta~\cite{poggi2005greta} are such examples, however, they do not support generative AI models such as Large Language Models (LLMs). Moreover, a recent wave of lightweight, high performing models (Speech-To-Text (STT)~\cite{rizk2019evaluation, peinl2020open}, LLM~\cite{abdin2024phi}, and TTS~\cite{casanova2024xtts}) can be run on-edge to overcome privacy concerns evident in cloud-based services. These gaps and improvements in current tools add additional motivation for the creation of a cohesive framework tailored for SIAs. 

Recent projects such as Pipecat~\cite{Flaqué2024pipecat} and NVIDIA ACE~\cite{nvidia-ace} provide frameworks for building conversational agents by connecting AI microservices. However, Pipecat is not tailored for building SIAs and lacks integration with game engines, which SIA research heavily relies upon. NVIDIA ACE, at the time of writing, is not open-source and restricts users to only running NVIDIA-approved AI microservices. This may be a critical limitation for research involving custom AI microservices and pipelines. Furthermore, NVIDIA ACE requires a cost-prohibitive enterprise plan if developers would like to host their microservices off-cloud or on-prem.

% It provides modularity but lacks essential support for building SIAs such as interfacing with any game engine or expansive front-end client, constraining its capabilities to a Python-based server.

% Our Contributions
\subsection{Limitations of Current Approaches.}
Currently, several factors hinder SIA research: 1) computational limitations of devices (e.g., head-mounted displays (HMDs)) which prohibit running advanced AI models on a standalone device, 2) hardware architecture incompatibilities with AI models which necessitate an inferencing server, and 3) high latency of cloud-based microservices. To address these shortcomings, we developed \textbf{Estuary}, a distributed framework for low-latency real-time SIAs. Estuary simplifies the development process by seamlessly integrating user-defined modules for ASR, TTS, dialogue management, and LLMs with the purpose of constructing pipelines for SIAs. This allows it to support low-latency real-time voice interactions, understand the physical world, and to facilitate reproducible and highly configurable experiments.

\begin{figure}
      \centering
      \includegraphics[width=0.4\textwidth]{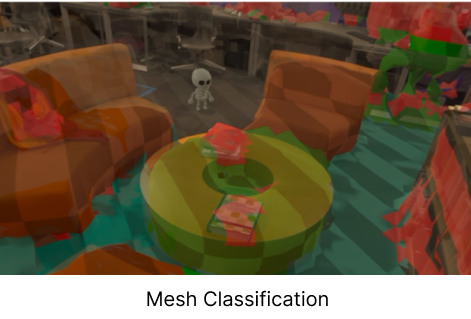}
      % \vspace{-14px}
      \caption{Estuary leveraging mesh classification capabilities into multimodal interaction.}
      \label{fig:teaser}
\end{figure}

\raggedbottom
\section{System}
Estuary brings five core value-propositions: 1) an interoperable microservice architecture, 2) multi-platform support, 3) off-cloud capabilities, 4) support for multimodal input and analysis, and 5) an open-source nature that opens it to community contributions. 

\subsection{Microservice Architecture.}
\begin{figure} [htbp]
    \centering
    \includegraphics[width=.4\textwidth]{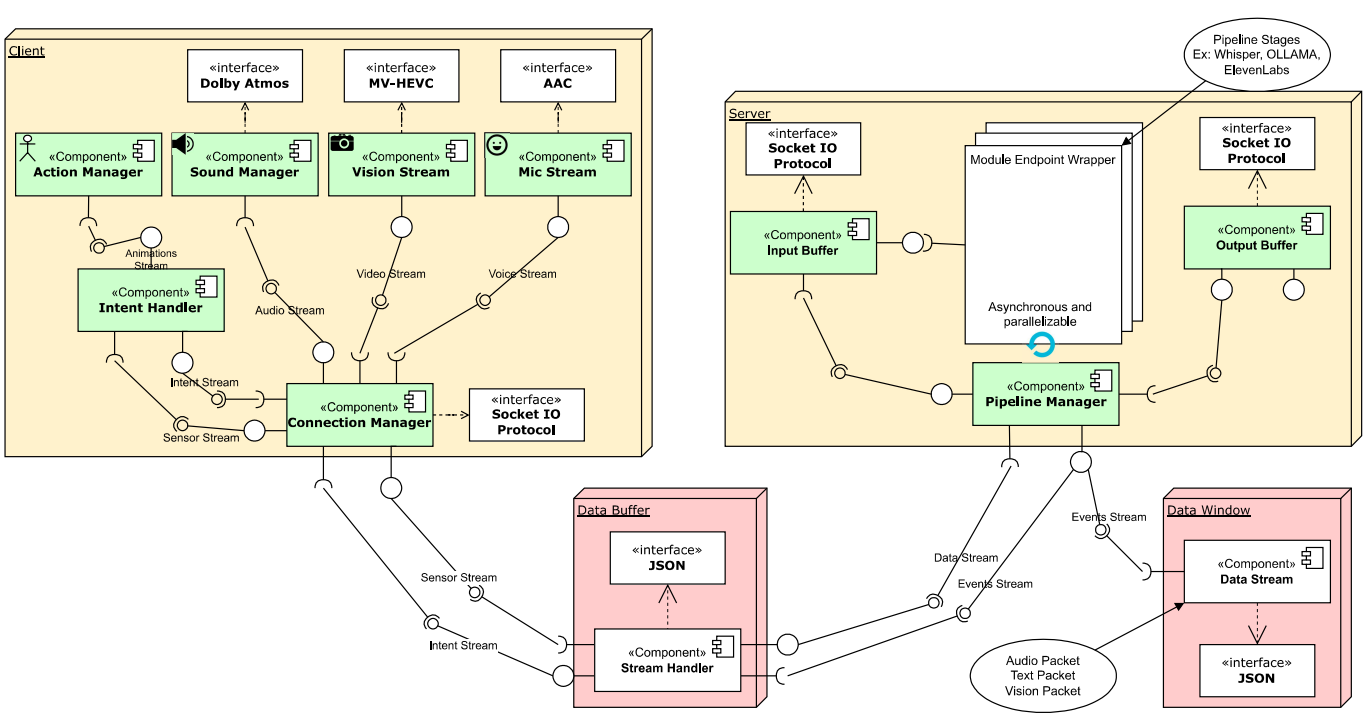}
    \caption{System diagram of the server and client}
    \label{fig:SystemDiagram}
    \vspace{-10px}
\end{figure}

We use a modular design to universally wrap local model or online API service within a \textit{Stage}, which are asynchronous and parallelizable as denoted in Figure \ref{fig:SystemDiagram}. A \textit{Stage} component wraps the ML inference logic and hosts it on a child process, aggregating inputs and dispatching outputs. A selection of \textit{Stages} (e.g., Whisper~\cite{radford2023robust}, GPT-3.5~\cite{gpt3.5}, gTTS~\cite{durette2024}) are connected into a $Pipeline$ according to the flow of choice. The $Pipeline$ internally orchestrates the flow of a standardized data type $DataWindow$, which consists of one or more $DataPacket$(s) (e.g., $AudioPacket$, $VisionPacket$, $TextPacket$, etc.). A $DataPacket$, with its source and creation timestamp, acts as identifiable placeholders for the \textit{Stage} outcomes. This micro-service architecture allows us to plug and play new models rapidly by implementing them as \textit{Stage}(s). A $Pipeline$ out of the box runs as a background process on a server that can communicate with any client through SocketIO protocol~\cite{socketio}. 

% --------------------
    
\subsection{Multi-platform.}
Estuary is a distributed framework that uses the SocketIO protocol to establish a connection between a client device and a host device running our framework. It can support any platform, including those supported by Unity game engine given that the hardware has a microphone, speaker, and can communicate using SocketIO. This hardware-agnostic support makes available HMDs and other devices by overcoming compute limitations or hardware architecture incompatibilities. 

\subsection{Off-Cloud.}
In addition to leveraging cloud APIs, Estuary can be hosted locally and/or entirely with off-cloud microservices. This reduces latency by eliminating multiple cloud endpoints, improves reproducibility, and adds privacy and security as no data ever enters the cloud. As shown in our video demo, it takes approximately 1.2$\sim$2.5 seconds (compared to an average 2.8 second latency from ChatGPT-4o ~\cite{HelloGPT-4o}) from the end of a user's dialogue to the first utterance from the SIA's TTS module through a FasterWhisperBase.EN~\cite{fasterwhisper} $\rightarrow$ GPT-3.5 API~\cite{gpt3.5} $\rightarrow$ XTTS~\cite{casanova2024xtts} pipeline on a desktop with a RTX 4090 graphics card. This is made possible through several optimizations such as simultaneously streaming the LLM and TTS response. Furthermore, off-cloud microservices ensure reproducibility, which is of utmost importance especially in fields relating to psychology~\cite{open2015estimating}. In Estuary, the exact same versions of LLM, ASR, TTS, and other microservices can be maintained and loaded for future use, whereas cloud-based services do not have guarantees to remain unchanged overtime. 

\subsection{Multimodal Input And Analysis.}
Multimodal input ~\cite{HelloGPT-4o} is critical for agents to understand the physical world and multiple modalities to produce a better cognitive model~\cite{2018openface, cao2019openpose}. Estuary integrates with Unity and packages like ARKit ~\cite{UnityARKitDocs} to empower embodied agents with basic semantic understanding of the physical world and pathfinding capabilities in Augmented Reality (AR). Currently, Estuary supports text and audio datastreams. We plan to expand to video as well which will shift reliance away from hardware-specific packages.

\subsection{Open-Source.}
Estuary is open-source and promotes growth from community contribution. By nature, Estuary is flexible and hugely extensible to support integration of microservices now and into the future. Researchers and developers have the freedom to choose what models they use and what data is collected from anywhere in the pipeline, all without limitations imposed by paid services. Estuary is built as a robust, modern framework that can empower SIA research for years to come.

% ======================================================

\section{Demonstration}
\label{section:demo}
Our first scenario consists of a computer hosting Estuary interfaced to an HMD over a local network to demonstrate an advanced embodied conversational agent that can classify objects in its physical surroundings and interact accordingly. The second scenario consists of a computer hosting both Estuary and a desktop frontend to demonstrate the versatility of our framework. Estuary's source code can be found in our GitHub~\footnote{\url{https://github.com/Al-Estuary}} and a demo video here on our website~\footnote{\url{https://estuary-ai.github.io/}}.

%%
%% The acknowledgments section is defined using the "acks" environment
%% (and NOT an unnumbered section). This ensures the proper
%% identification of the section in the article metadata, and the
%% consistent spelling of the heading.
\section{Acknowledgments}
We would like to thank our teammates and advisors from the 2022 and 2023 NASA SUITS Challenge whose efforts contributed to the success of this project. We would also like to thank all open-source developers and creatives whose work we respectfully used.

%%
%% The next two lines define the bibliography style to be used, and
%% the bibliography file.
% \bibliographystyle{ACM-Reference-Format}
% \bibliography{references}

%%% -*-BibTeX-*-
%%% Do NOT edit. File created by BibTeX with style
%%% ACM-Reference-Format-Journals [18-Jan-2012].

\begin{thebibliography}{26}



%%% ====================================================================
%%% NOTE TO THE USER: you can override these defaults by providing
%%% customized versions of any of these macros before the \bibliography
%%% command.  Each of them MUST provide its own final punctuation,
%%% except for \shownote{}, \showDOI{}, and \showURL{}.  The latter two
%%% do not use final punctuation, in order to avoid confusing it with
%%% the Web address.
%%%
%%% To suppress output of a particular field, define its macro to expand
%%% to an empty string, or better, \unskip, like this:
%%%
%%% \newcommand{\showDOI}[1]{\unskip}   % LaTeX syntax
%%%
%%% \def \showDOI #1{\unskip}           % plain TeX syntax
%%%
%%% ====================================================================

\ifx \showCODEN    \undefined \def \showCODEN     #1{\unskip}     \fi
\ifx \showDOI      \undefined \def \showDOI       #1{#1}\fi
\ifx \showISBNx    \undefined \def \showISBNx     #1{\unskip}     \fi
\ifx \showISBNxiii \undefined \def \showISBNxiii  #1{\unskip}     \fi
\ifx \showISSN     \undefined \def \showISSN      #1{\unskip}     \fi
\ifx \showLCCN     \undefined \def \showLCCN      #1{\unskip}     \fi
\ifx \shownote     \undefined \def \shownote      #1{#1}          \fi
\ifx \showarticletitle \undefined \def \showarticletitle #1{#1}   \fi
\ifx \showURL      \undefined \def \showURL       {\relax}        \fi
% The following commands are used for tagged output and should be
% invisible to TeX
\providecommand\bibfield[2]{#2}
\providecommand\bibinfo[2]{#2}
\providecommand\natexlab[1]{#1}
\providecommand\showeprint[2][]{arXiv:#2}

\bibitem[fas({[n.\,d.]})]%
        {fasterwhisper}
 \bibinfo{year}{[n.\,d.]}\natexlab{}.
\newblock \bibinfo{title}{Faster Whisper}.
\newblock \bibinfo{howpublished}{\url{https://github.com/SYSTRAN/faster-whisper?tab=readme-ov-file}}.
\newblock


\bibitem[nvi({[n.\,d.]})]%
        {nvidia-ace}
 \bibinfo{year}{[n.\,d.]}\natexlab{}.
\newblock \bibinfo{title}{NVIDIA Ace}.
\newblock \bibinfo{howpublished}{\url{https://developer.nvidia.com/ace}}.
\newblock


\bibitem[soc({[n.\,d.]})]%
        {socketio}
 \bibinfo{year}{[n.\,d.]}\natexlab{}.
\newblock \bibinfo{title}{Socket.IO}.
\newblock
\newblock
\urldef\tempurl%
\url{https://socket.io/}
\showURL{%
\tempurl}


\bibitem[Abdin et~al\mbox{.}(2024)]%
        {abdin2024phi}
\bibfield{author}{\bibinfo{person}{Marah Abdin}, \bibinfo{person}{Sam~Ade Jacobs}, \bibinfo{person}{Ammar~Ahmad Awan}, \bibinfo{person}{Jyoti Aneja}, \bibinfo{person}{Ahmed Awadallah}, \bibinfo{person}{Hany Awadalla}, \bibinfo{person}{Nguyen Bach}, \bibinfo{person}{Amit Bahree}, \bibinfo{person}{Arash Bakhtiari}, \bibinfo{person}{Harkirat Behl}, {et~al\mbox{.}}} \bibinfo{year}{2024}\natexlab{}.
\newblock \showarticletitle{Phi-3 technical report: A highly capable language model locally on your phone}.
\newblock \bibinfo{journal}{\emph{arXiv preprint arXiv:2404.14219}} (\bibinfo{year}{2024}).
\newblock


\bibitem[Baltrusaitis et~al\mbox{.}(2018)]%
        {2018openface}
\bibfield{author}{\bibinfo{person}{Tadas Baltrusaitis}, \bibinfo{person}{Amir Zadeh}, \bibinfo{person}{Yao~Chong Lim}, {and} \bibinfo{person}{Louis-Philippe Morency}.} \bibinfo{year}{2018}\natexlab{}.
\newblock \showarticletitle{OpenFace 2.0: Facial Behavior Analysis Toolkit}. In \bibinfo{booktitle}{\emph{2018 13th IEEE International Conference on Automatic Face and Gesture Recognition (FG 2018)}}. \bibinfo{pages}{59--66}.
\newblock
\urldef\tempurl%
\url{https://doi.org/10.1109/FG.2018.00019}
\showDOI{\tempurl}


\bibitem[Brown et~al\mbox{.}(2020)]%
        {gpt3.5}
\bibfield{author}{\bibinfo{person}{Tom~B. Brown}, \bibinfo{person}{Benjamin Mann}, \bibinfo{person}{Nick Ryder}, \bibinfo{person}{Melanie Subbiah}, \bibinfo{person}{Jared Kaplan}, \bibinfo{person}{Prafulla Dhariwal}, \bibinfo{person}{Arvind Neelakantan}, \bibinfo{person}{Pranav Shyam}, \bibinfo{person}{Girish Sastry}, \bibinfo{person}{Amanda Askell}, \bibinfo{person}{Sandhini Agarwal}, \bibinfo{person}{Ariel Herbert{-}Voss}, \bibinfo{person}{Gretchen Krueger}, \bibinfo{person}{Tom Henighan}, \bibinfo{person}{Rewon Child}, \bibinfo{person}{Aditya Ramesh}, \bibinfo{person}{Daniel~M. Ziegler}, \bibinfo{person}{Jeffrey Wu}, \bibinfo{person}{Clemens Winter}, \bibinfo{person}{Christopher Hesse}, \bibinfo{person}{Mark Chen}, \bibinfo{person}{Eric Sigler}, \bibinfo{person}{Mateusz Litwin}, \bibinfo{person}{Scott Gray}, \bibinfo{person}{Benjamin Chess}, \bibinfo{person}{Jack Clark}, \bibinfo{person}{Christopher Berner}, \bibinfo{person}{Sam McCandlish}, \bibinfo{person}{Alec Radford}, \bibinfo{person}{Ilya Sutskever},
  {and} \bibinfo{person}{Dario Amodei}.} \bibinfo{year}{2020}\natexlab{}.
\newblock \showarticletitle{Language Models are Few-Shot Learners}.
\newblock \bibinfo{journal}{\emph{CoRR}}  \bibinfo{volume}{abs/2005.14165} (\bibinfo{year}{2020}).
\newblock
\showeprint[arXiv]{2005.14165}
\urldef\tempurl%
\url{https://arxiv.org/abs/2005.14165}
\showURL{%
\tempurl}


\bibitem[Cao et~al\mbox{.}(2019)]%
        {cao2019openpose}
\bibfield{author}{\bibinfo{person}{Zhe Cao}, \bibinfo{person}{Gines Hidalgo}, \bibinfo{person}{Tomas Simon}, \bibinfo{person}{Shih-En Wei}, {and} \bibinfo{person}{Yaser Sheikh}.} \bibinfo{year}{2019}\natexlab{}.
\newblock \bibinfo{title}{OpenPose: Realtime Multi-Person 2D Pose Estimation using Part Affinity Fields}.
\newblock
\newblock
\showeprint[arxiv]{1812.08008}


\bibitem[Casanova et~al\mbox{.}(2024)]%
        {casanova2024xtts}
\bibfield{author}{\bibinfo{person}{Edresson Casanova}, \bibinfo{person}{Kelly Davis}, \bibinfo{person}{Eren Gölge}, \bibinfo{person}{Görkem Göknar}, \bibinfo{person}{Iulian Gulea}, \bibinfo{person}{Logan Hart}, \bibinfo{person}{Aya Aljafari}, \bibinfo{person}{Joshua Meyer}, \bibinfo{person}{Reuben Morais}, \bibinfo{person}{Samuel Olayemi}, {and} \bibinfo{person}{Julian Weber}.} \bibinfo{year}{2024}\natexlab{}.
\newblock \bibinfo{title}{XTTS: a Massively Multilingual Zero-Shot Text-to-Speech Model}.
\newblock
\newblock
\showeprint[arxiv]{2406.04904}~[eess.AS]


\bibitem[Collaboration(2015)]%
        {open2015estimating}
\bibfield{author}{\bibinfo{person}{Open~Science Collaboration}.} \bibinfo{year}{2015}\natexlab{}.
\newblock \showarticletitle{Estimating the reproducibility of psychological science}.
\newblock \bibinfo{journal}{\emph{Science}} \bibinfo{volume}{349}, \bibinfo{number}{6251} (\bibinfo{year}{2015}), \bibinfo{pages}{aac4716}.
\newblock


\bibitem[Durette(2024)]%
        {durette2024}
\bibfield{author}{\bibinfo{person}{Pierre~Nicolas Durette}.} \bibinfo{year}{2024}\natexlab{}.
\newblock \bibinfo{title}{gTTS}.
\newblock \bibinfo{howpublished}{\url{https://github.com/pndurette/gTTS}}.
\newblock


\bibitem[Flaqué et~al\mbox{.}(2024)]%
        {Flaqué2024pipecat}
\bibfield{author}{\bibinfo{person}{Aleix~Conchillo Flaqué}, \bibinfo{person}{Moishe Lettvin}, \bibinfo{person}{Kwindla~Hultman Kramer}, \bibinfo{person}{chadbailey59}, \bibinfo{person}{Jon Taylor}, \bibinfo{person}{Thomas B.}, \bibinfo{person}{Liza}, \bibinfo{person}{James Hush}, {and} \bibinfo{person}{Rahul Nair}.} \bibinfo{year}{2024}\natexlab{}.
\newblock \bibinfo{booktitle}{\emph{pipecat-ai/pipecat}}.
\newblock
\urldef\tempurl%
\url{https://github.com/pipecat-ai/pipecat}
\showURL{%
\tempurl}


\bibitem[Gao et~al\mbox{.}(2023)]%
        {gao2023s}
\bibfield{author}{\bibinfo{person}{Chen Gao}, \bibinfo{person}{Xiaochong Lan}, \bibinfo{person}{Zhihong Lu}, \bibinfo{person}{Jinzhu Mao}, \bibinfo{person}{Jinghua Piao}, \bibinfo{person}{Huandong Wang}, \bibinfo{person}{Depeng Jin}, {and} \bibinfo{person}{Yong Li}.} \bibinfo{year}{2023}\natexlab{}.
\newblock \showarticletitle{S3: Social-network Simulation System with Large Language Model-Empowered Agents}.
\newblock \bibinfo{journal}{\emph{arXiv preprint arXiv:2307.14984}} (\bibinfo{year}{2023}).
\newblock


\bibitem[Hartholt et~al\mbox{.}(2013)]%
        {hartholt2013all}
\bibfield{author}{\bibinfo{person}{Arno Hartholt}, \bibinfo{person}{David Traum}, \bibinfo{person}{Stacy~C Marsella}, \bibinfo{person}{Ari Shapiro}, \bibinfo{person}{Giota Stratou}, \bibinfo{person}{Anton Leuski}, \bibinfo{person}{Louis-Philippe Morency}, {and} \bibinfo{person}{Jonathan Gratch}.} \bibinfo{year}{2013}\natexlab{}.
\newblock \showarticletitle{All together now: Introducing the virtual human toolkit}. In \bibinfo{booktitle}{\emph{Intelligent Virtual Agents: 13th International Conference, IVA 2013, Edinburgh, UK, August 29-31, 2013. Proceedings 13}}. Springer, \bibinfo{pages}{368--381}.
\newblock


\bibitem[Hasan et~al\mbox{.}(2023)]%
        {hasan2023sapien}
\bibfield{author}{\bibinfo{person}{Masum Hasan}, \bibinfo{person}{Cengiz Ozel}, \bibinfo{person}{Sammy Potter}, {and} \bibinfo{person}{Ehsan Hoque}.} \bibinfo{year}{2023}\natexlab{}.
\newblock \showarticletitle{SAPIEN: affective virtual agents powered by large language models}. In \bibinfo{booktitle}{\emph{2023 11th International Conference on Affective Computing and Intelligent Interaction Workshops and Demos (ACIIW)}}. IEEE, \bibinfo{pages}{1--3}.
\newblock


\bibitem[Li et~al\mbox{.}({[n.\,d.]})]%
        {lieconagent}
\bibfield{author}{\bibinfo{person}{Nian Li}, \bibinfo{person}{Chen Gao}, \bibinfo{person}{Mingyu Li}, \bibinfo{person}{Yong Li}, {and} \bibinfo{person}{Qingmin Liao}.} \bibinfo{year}{[n.\,d.]}\natexlab{}.
\newblock \showarticletitle{EconAgent: Large Language Model-Empowered Agents for Simulating Macroeconomic Activities}.
\newblock  (\bibinfo{year}{[n.\,d.]}).
\newblock


\bibitem[Liang et~al\mbox{.}(2023)]%
        {liang2023encouraging}
\bibfield{author}{\bibinfo{person}{Tian Liang}, \bibinfo{person}{Zhiwei He}, \bibinfo{person}{Wenxiang Jiao}, \bibinfo{person}{Xing Wang}, \bibinfo{person}{Yan Wang}, \bibinfo{person}{Rui Wang}, \bibinfo{person}{Yujiu Yang}, \bibinfo{person}{Zhaopeng Tu}, {and} \bibinfo{person}{Shuming Shi}.} \bibinfo{year}{2023}\natexlab{}.
\newblock \showarticletitle{Encouraging divergent thinking in large language models through multi-agent debate}.
\newblock \bibinfo{journal}{\emph{arXiv preprint arXiv:2305.19118}} (\bibinfo{year}{2023}).
\newblock


\bibitem[Lugrin et~al\mbox{.}(2022)]%
        {lugrin2022handbook}
\bibfield{author}{\bibinfo{person}{Birgit Lugrin}, \bibinfo{person}{Catherine Pelachaud}, {and} \bibinfo{person}{David Traum}.} \bibinfo{year}{2022}\natexlab{}.
\newblock \bibinfo{booktitle}{\emph{The Handbook on Socially Interactive Agents: 20 Years of Research on Embodied Conversational Agents, Intelligent Virtual Agents, and Social Robotics Volume 2: Interactivity, Platforms, Application}}.
\newblock \bibinfo{publisher}{ACM}.
\newblock


\bibitem[OpenAI(2024)]%
        {HelloGPT-4o}
\bibfield{author}{\bibinfo{person}{OpenAI}.} \bibinfo{year}{2024}\natexlab{}.
\newblock \bibinfo{title}{Hello GPT-4o}.
\newblock
\newblock
\urldef\tempurl%
\url{https://openai.com/index/hello-gpt-4o/}
\showURL{%
\tempurl}


\bibitem[Park et~al\mbox{.}(2023)]%
        {park2023generative}
\bibfield{author}{\bibinfo{person}{Joon~Sung Park}, \bibinfo{person}{Joseph O'Brien}, \bibinfo{person}{Carrie~Jun Cai}, \bibinfo{person}{Meredith~Ringel Morris}, \bibinfo{person}{Percy Liang}, {and} \bibinfo{person}{Michael~S Bernstein}.} \bibinfo{year}{2023}\natexlab{}.
\newblock \showarticletitle{Generative agents: Interactive simulacra of human behavior}. In \bibinfo{booktitle}{\emph{Proceedings of the 36th Annual ACM Symposium on User Interface Software and Technology}}. \bibinfo{pages}{1--22}.
\newblock


\bibitem[Peinl et~al\mbox{.}(2020)]%
        {peinl2020open}
\bibfield{author}{\bibinfo{person}{Ren{\'e} Peinl}, \bibinfo{person}{Basem Rizk}, {and} \bibinfo{person}{Robert Szabad}.} \bibinfo{year}{2020}\natexlab{}.
\newblock \showarticletitle{Open source speech recognition on edge devices}. In \bibinfo{booktitle}{\emph{2020 10th International Conference on Advanced Computer Information Technologies (ACIT)}}. IEEE, \bibinfo{pages}{441--445}.
\newblock


\bibitem[Poggi et~al\mbox{.}(2005)]%
        {poggi2005greta}
\bibfield{author}{\bibinfo{person}{Isabella Poggi}, \bibinfo{person}{Catherine Pelachaud}, \bibinfo{person}{Fiorella de Rosis}, \bibinfo{person}{Valeria Carofiglio}, {and} \bibinfo{person}{Berardina De~Carolis}.} \bibinfo{year}{2005}\natexlab{}.
\newblock \showarticletitle{Greta. a believable embodied conversational agent}.
\newblock In \bibinfo{booktitle}{\emph{Multimodal intelligent information presentation}}. \bibinfo{publisher}{Springer}, \bibinfo{pages}{3--25}.
\newblock


\bibitem[Radford et~al\mbox{.}(2023)]%
        {radford2023robust}
\bibfield{author}{\bibinfo{person}{Alec Radford}, \bibinfo{person}{Jong~Wook Kim}, \bibinfo{person}{Tao Xu}, \bibinfo{person}{Greg Brockman}, \bibinfo{person}{Christine McLeavey}, {and} \bibinfo{person}{Ilya Sutskever}.} \bibinfo{year}{2023}\natexlab{}.
\newblock \showarticletitle{Robust speech recognition via large-scale weak supervision}. In \bibinfo{booktitle}{\emph{International Conference on Machine Learning}}. PMLR, \bibinfo{pages}{28492--28518}.
\newblock


\bibitem[Rizk(2019)]%
        {rizk2019evaluation}
\bibfield{author}{\bibinfo{person}{Basem Rizk}.} \bibinfo{year}{2019}\natexlab{}.
\newblock \showarticletitle{Evaluation of state of art open-source ASR engines with local inferencing}.
\newblock In \bibinfo{booktitle}{\emph{Evaluation of State Of Art Open-source ASR Engines with Local Inferencing}}. Vol.~\bibinfo{volume}{8}.
\newblock


\bibitem[Shaikh et~al\mbox{.}(2024)]%
        {shaikh2024rehearsal}
\bibfield{author}{\bibinfo{person}{Omar Shaikh}, \bibinfo{person}{Valentino~Emil Chai}, \bibinfo{person}{Michele Gelfand}, \bibinfo{person}{Diyi Yang}, {and} \bibinfo{person}{Michael~S Bernstein}.} \bibinfo{year}{2024}\natexlab{}.
\newblock \showarticletitle{Rehearsal: Simulating conflict to teach conflict resolution}. In \bibinfo{booktitle}{\emph{Proceedings of the CHI Conference on Human Factors in Computing Systems}}. \bibinfo{pages}{1--20}.
\newblock


\bibitem[Unity(2024)]%
        {UnityARKitDocs}
\bibfield{author}{\bibinfo{person}{Unity}.} \bibinfo{year}{2024}\natexlab{}.
\newblock \bibinfo{title}{UnityARKitDocumentation}.
\newblock
\newblock
\urldef\tempurl%
\url{https://docs.unity3d.com/Packages/com.unity.xr.arkit@5.1/manual/}
\showURL{%
\tempurl}


\bibitem[Wang et~al\mbox{.}(2023)]%
        {wang2023humanoid}
\bibfield{author}{\bibinfo{person}{Zhilin Wang}, \bibinfo{person}{Yu~Ying Chiu}, {and} \bibinfo{person}{Yu~Cheung Chiu}.} \bibinfo{year}{2023}\natexlab{}.
\newblock \showarticletitle{Humanoid agents: Platform for simulating human-like generative agents}.
\newblock \bibinfo{journal}{\emph{arXiv preprint arXiv:2310.05418}} (\bibinfo{year}{2023}).
\newblock


\end{thebibliography}
%%% -*-BibTeX-*-
%%% Do NOT edit. File created by BibTeX with style
%%% ACM-Reference-Format-Journals [18-Jan-2012].

%%
%% If your work has an appendix, this is the place to put it.
\appendix
% \section{Technical Requirements}

% \noindent Apple Vision Pro demonstration: 
% \begin{itemize}
%     \item 1 RTX 4090 or 24GB/higher VRAM equivalent desktop 
%     \item 1 Apple Vision Pro headset
%     \item 2 chairs for demonstration purposes
%     \item a fast network connection that also allows SocketIO communications between devices
%     \item 1 external monitor and HDMI cord
%     \item 1 table
% \end{itemize}

% \noindent Second Demonstration:
% \begin{itemize}
%     \item 1 RTX 4090 or 24GB/higher VRAM equivalent desktop 
%     \item 1 external monitor and HDMI cord
%     \item 1 table
% \end{itemize}

\end{document}